\documentclass[11pt]{article}
\usepackage{graphicx}

\newcommand{\BABARPubYear}    {02}

\newcommand{\BABARConfNumber} {06}
\newcommand{\SLACPubNumber} {9166}

\input babarsym

\long\def\inst#1{\par\nobreak\kern 4pt\nobreak
    {\it #1}\par\vskip 10pt plus 3pt minus 3pt}

\begin{document}
{\pagestyle{empty}

\begin{flushright}
\babar-CONF-\BABARPubYear/\BABARConfNumber \\
SLAC-PUB-\SLACPubNumber \\
\end{flushright}

\par\vskip 5cm

\begin{center}
\Large \bf \boldmath
Rare $B$ Decays to States Containing a $J/\psi$ Meson
\end{center}
\bigskip

\begin{center}
\large The \babar\ Collaboration\\
\mbox{ }\\
\today
\end{center}
\bigskip \bigskip

\begin{center}
\large \bf Abstract
\end{center}
We report preliminary measurements of the branching fractions for 
$B^+\rightarrow J/\psi \phi K^+$, $B^0\rightarrow J/\psi \phi K_S^0$, $B^0\rightarrow J/\psi \phi$, $B^0\rightarrow J/\psi \eta$ and $B^0\rightarrow J/\psi \eta^\prime$ using {56} million $B\bar B$ events collected 
at the $\FourS$ resonance
with the $\babar$ detector at PEP-II. We measure branching fractions of ${\cal B}$($B^+\rightarrow J/\psi \phi K^{+}$)=
$(4.4\pm 1.4(stat) \pm 0.7(syst))$$\times 10^{-5}$ and ${\cal B}$($B^0\rightarrow J/\psi
\phi K_{S}^{0}$)=$(5.1\pm 1.9(stat) \pm 0.9(syst))$$\times 10^{-5}$, and set upper
limits at 90$\%$ C.L. for branching fractions ${\cal B}$($B^0\rightarrow J/\psi \phi$)$< 0.95\times 10^{-5}$, ${\cal B}$($B^0\rightarrow J/\psi \eta $)$< 2.7\times 10^{-5}$, and ${\cal B}$($B^0\rightarrow J/\psi \eta ^{\prime }$)$< 6.4\times 10^{-5}$.

\vfill
\begin{center}
Presented at the XXXVII$^{th}$ Rencontres de Moriond on
QCD and Hadronic Interactions,\\
3-16---3/23/2002, Les Arcs, Savoie, France
\end{center}

\vspace{1.0cm}
\begin{center}
{\em Stanford Linear Accelerator Center, Stanford University, 
Stanford, CA 94309} \\ \vspace{0.1cm}\hrule\vspace{0.1cm}
Work supported in part by Department of Energy contract DE-AC03-76SF00515.
\end{center}

}

\begin{center}
\small

The \babar\ Collaboration,
\bigskip

B.~Aubert,
D.~Boutigny,
J.-M.~Gaillard,
A.~Hicheur,
Y.~Karyotakis,
J.~P.~Lees,
P.~Robbe,
V.~Tisserand,
A.~Zghiche
\inst{Laboratoire de Physique des Particules, F-74941 Annecy-le-Vieux, France }
A.~Palano,
A.~Pompili
\inst{Universit\`a di Bari, Dipartimento di Fisica and INFN, I-70126 Bari, Italy }
% ...

\end{center}\newpage

\setcounter{footnote}{0}

\section{\protect\smallskip Introduction}

\begin{figure}[!htb]
\begin{center}
\vspace{-1.0truein}
\includegraphics[height=7cm]{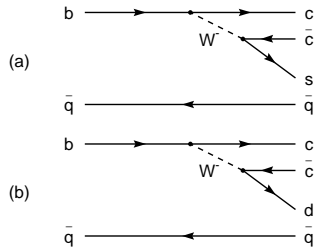}
\caption{
Quark diagrams for (a) $%
B\rightarrow J/\psi K$ and $J/\psi K^{*}$ and (b) $B\rightarrow J/\psi \pi $
and $J/\psi \rho .$
}
\label{fig:feyn-jpsik}
\end{center}
\end{figure}

\begin{figure}[!htb]
\begin{center}
\includegraphics[height=7cm]{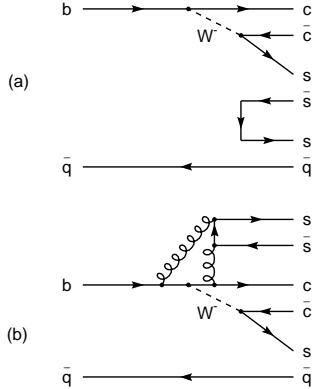}
\caption{
Quark diagrams for $B\rightarrow J/\psi \phi K$ via (a) strange sea quarks and (b) gluon
coupling.
}
\label{fig:feyn-jpsiphik}
\end{center}
\end{figure}

The Cabibbo-favored transition $b\rightarrow c\overline{c}s$ is well
established by observation~\cite{babar-charmonium} 
of $B$ decays to a charmonium state and a kaon, such as $B\rightarrow J/\psi K$ and $J/\psi K^{*}$. Recent observations of the decays $B\rightarrow J/\psi \pi$~\cite{babar-charmonium} and $J/\psi \rho~\cite{babar-jpsirho}$ are evidence for the Cabibbo-suppressed transition $b\rightarrow c\overline{c}d$. The quark diagrams for these color-suppressed decays are shown in Figures~\ref{fig:feyn-jpsik} (a) and (b). We search for $B$ meson decays into other final states with charmonium. Since $B\rightarrow J/\psi \pi $ is observed, the
Cabibbo suppressed modes $J/\psi \eta $ and $J/\psi \eta ^{\prime }$ may
exist at a comparable level. A further test is to search for quark
combinations such as $b\overline{q}\rightarrow c\overline{c}s\overline{s}s
\overline{q}$, where the $s\overline{s}$ quark pairs are produced from
sea quarks or are connected via external
gluons as shown in Figures~\ref{fig:feyn-jpsiphik} (a) and (b). This would be
exemplified in modes such as $B\rightarrow J/\psi \phi K$. The mode $J/\psi \phi $ is a pure rescattering process, the measurement of which can help to resolve the discrete ambiguity in the $cos(2\beta)$ measurement with $B\rightarrow J/\psi K^{*}$~\cite{resc}. In this paper
we report a search for $B$ decays into $J/\psi \phi ,$ $J/\psi \eta $, $
J/\psi \eta ^{\prime }$, $J/\psi \phi K^{+}$, and $J/\psi \phi K_{S}^{0}$
and present their branching fractions or upper limits.

Using a factorization approximation with heavy quarks, A. Deandrea\textit{
et al.}~\cite{deandrea} have predicted the branching fraction of $B\rightarrow $ $
J/\psi \eta $ to be a factor of 3.7 smaller than $B^{0}\rightarrow $ $J/\psi
\pi ^{0}$, corresponding to ${\cal B}(B^{0}\rightarrow $ $J/\psi \eta )=0.54\times 10^{-5}~\cite
{jpsi-eta-br}$. The L3 Collaboration~\cite{L3} searched for this mode, found
no events and set an upper limit, ${\cal B}(B^{0}\rightarrow $ $J/\psi \eta)<1.2\times 10^{-3}$ at 90\% confidence level. The mode $B\rightarrow J/\psi \phi K$ has
been observed by the CLEO Collaboration~\cite{cleo} with 10 events in $9.6\times 10^{6}$ $B\overline{B}$ pairs with the result ${\cal B}(B\rightarrow J/\psi \phi K)=\left( 8.8_{-3.0}^{+3.5}\pm 1.3\right) \times 10^{-5}$. In addition to yielding $c\overline{c}$ bound states, the decay $B\rightarrow X(c\overline{c})+K$ may provide hybrid charmonium ($c\overline{c}+glue$)~\cite{close}, and the hybrid state may ultimately decay into $J/\psi \phi $
in the final state $J/\psi \phi K$. No published results exist
for the modes $B\rightarrow J/\psi \phi $ and $J/\psi \eta ^{\prime }$.


\section{$\babar$ Detector and Dataset}

The data used in this analysis were collected with the $\babar$ detector at
the PEP-II asymmetric $e^{+}e^{-}$ storage ring. The complete detector is
described in detail elsewhere~\cite{babar-det}. We briefly describe the
relevant detector subsystems for the physics analysis in this paper. The $\babar$
detector contains a five-layer silicon vertex tracker (SVT) and a forty-layer drift chamber (DCH) in a 1.5-Tesla solenoidal magnetic field. These devices detect charged particles and measure their momentum and
energy loss. The transverse momentum resolution is $\sigma
_{p_{t}}/p_{t}=(0.13\pm 0.01)\times p_{t}\%+(0.45\pm 0.03)\%$, where $p_{t}$
is measured in GeV/$c$. Photons and neutral hadrons are detected in a
CsI(Tl) crystal electromagnetic calorimeter (EMC). The EMC detects
photons with energies as low as 20 MeV and identifies electrons by their
large energy deposit. The EMC energy resolution for photons and electrons
is $\sigma (E)/E=2.3\%/E$(GeV)$^{1/4}+1.9\%$. The charged particle
identification (PID) combines SVT and DCH track energy loss measurements and
particle velocity measurements by an internally reflecting ring-imaging
Cherenkov detector (DIRC) of quartz bars circumjacent to the DCH. The
slotted steel flux return is instrumented with 18-19 layers of
planar resistive plate chambers (IFR). The IFR identifies penetrating
muons and neutral hadrons.

The data used in these analyses were collected in two periods, October 1999 to
October 2000 and February 2001 to December 2001.
The data correspond to a total integrated luminosity of 
$\sim 51$ fb$^{-1}$ taken on the $\FourS$
resonance and {6.3} fb$^{-1}$ taken off-resonance at an energy
0.04 GeV below the $\FourS$ center of mass energy and
below the threshold for $B\bar{B}$ production. This data set contains 
$\sim 56$ million $B\bar{B}$ events ($N_{B
\overline{B}}$).

\section{Physics Analysis}

\subsection{Particle Selection}
This analysis begins with selection of charged particles and photons.
All charged particle track candidates must have at least 12 DCH hits and $p_{t}>$
100 MeV/$c$. The track candidates not associated with a $K^0_{s}$ decay must also
extrapolate to a nominal interaction point within $\sqrt{x^{2}+y^{2}}<$1.5
c\.{m} and $\left| z\right| <$3 cm where the origin is at the interaction point, the $z$ axis is along the electron beam direction, the $y$ axis is vertically up, and the $x$ 
axis points away from the collider center. 
The muon, electron, and kaon candidates must have a polar angle in
radians of $0.3<\theta _{\mu }<2.7$, $0.41<\theta _{e}<2.409$, and $
0.45<\theta _{K}<2.5$, respectively.
In addition, all charged kaon candidates used in this analysis are required
to a lab momentum greater than 250 MeV/$c$.
These restrictions keep the tracks in regions that are
well understood by the PID systems.

Photons candidates are identified as hits in contiguous EMC crystals that are
summed together to form shower clusters and have a minimum 30 MeV
shower energy and satisfy certain shower shape criteria expected for
electromagnetic showers. The variables that describe the shower shape
include the lateral energy~\cite{LAT} (LAT) that determines the radial
energy profile, and Zernike moment~\cite{zernike} ($A_{42}$) that measures the
asymmetry of the cluster shape about its maximum. For electron showers the LAT peaks near 0.25 and $A_{42}$ peaks near zero. All the photon candidates
are required to have LAT$<0.8$.

Electron candidates are required to have a good match 
between the expected and measured energy loss (d$E$/d$x$) and 
between the expected and measured DIRC Cherenkov angle ($\theta _{C}$). 
Also the measurements of 
the ratio of EMC
shower energy over DCH momentum ($E/p$), and the number of EMC crystals associated with the track
candidate must be appropriate for an electron. We define very tight (VTE) and loose (LE) electron selection criteria that have efficiencies of 88\% and 97\%, respectively.

Muon candidates are required to have measurements of several variables that
help distinguish muons from other charged particles. These measurements are: the EMC energy, the number of hit layers in the IFR, the penetration depth expressed in units of interaction length along the track's path in the IFR and EMC, the difference between the expected and measured number of interaction lengths, the average number of hits per IFR layer, the variance of the distribution of the number of hits on each IFR layer, the fraction of hit layers between the innermost and outermost layer, the chi-square match of hits in the IFR, and the chi-square match between the IFR and the extrapolated DCH track.
We combine these variables to form different selection criteria applicable in different modes. The criteria are called tight (TM, efficiency 70\%), loose (LM, efficiency 86\%), and very loose (VLM,  efficiency 92\%).

Charged kaon candidates are selected based on d$E$/d$x$
information from the SVT and DCH and $\theta _{C}$. A
likelihood function that combines all the information is constructed for the
kaon, the proton and the pion hypotheses. A
likelihood ratio test determines if the candidate track satisfies the loose kaon
selection (LK), very tight kaon selection (VTK) or the not-a-pion selection (NP). The SVT, DCH and DIRC information and the likelihoods are used in certain selected
momentum ranges. 
The loose and very tight selections have typical efficiencies from 70  to 90\%, 
whereas
the extremely loose selection, not-a-pion, has $>90$\% efficiency.

\subsection{Event Selection}

The estimation of the signal and the background employs two kinematic
variables; the energy difference $\Delta E$, which is the energy of the $B$
candidate in the $\FourS$ frame minus the energy of the beam
particle, $E_{beam}^{CM}$, and the energy substituted mass $M_{ES}$ which
is $\sqrt{\left( E_{beam}^{CM}\right) -\left( P_{B}^{CM}\right) ^{2}}$, where 
$P_{B}^{CM}$ is the momentum of the $B$ candidate in the $\FourS$ frame.
Typically these two weakly correlated variables form a two dimensional
Gaussian distribution for the $B$ meson signal and a nearly flat two
dimensional distribution for background. The resolutions in $\Delta E$ and $M_{ES}$ can be different for different decay modes.

The intermediate state particles in this analysis are $J/\psi \left(
ee,\mu \mu \right) $, $\phi \left( K^{+}K^{-}\right) $, $\eta~(\gamma\gamma, \pi^{+}\pi ^{-}$$\pi^{0})$, $\eta ^{\prime }\left( \eta
\left( \gamma \gamma \right) \pi ^{+}\pi ^{-}\right) $, $\pi ^{0}\left(
\gamma \gamma \right) $,and $K_{S}^{0}\left( \pi ^{+}\pi ^{-}\right) $. All
of the intermediate state particles are selected in mass windows, which are listed in
Table~\ref{table-mass}. The $J/\psi \rightarrow ee$ decay has a slightly
asymmetric mass window to include the radiative $J/\psi $ tail. Since $B^{0}\rightarrow J/\psi \eta$ and $B^{0}\rightarrow J/\psi \eta^{\prime}$ are pseudoscalar decays into a vector and a pseudoscalar, the distribution of the helicity 
angle\footnote{
The lepton helicity angle is defined as the angle measured in the $J/\psi$ rest frame between the direction of the negative charged lepton and the direction opposite to the parent $B$ meson.} 
of the lepton, $\theta _{Lepton}$, from the $J/\psi$
is proportional to $\sin ^{2}\theta_{Lepton} $. Hence an additional cut of $
\left| \cos \theta _{Lepton}\right| <0.8$ is applied to reject continuum and other backgrounds. For the $\eta$ 
candidates, a $\pi^0$ veto is applied where the candidate is rejected
if either of the associated photons can be combined with another photon in
the event to form a $\gamma \gamma $ mass within 20 MeV/$c^{2}$ of the $\pi^0$ mass. Also for the mode $B^{0}\rightarrow J/\psi \eta(\gamma\gamma)$ the $\eta$ candidate is rejected for asymmetric
decays with $\left| \cos \theta _{\gamma }^{\eta }\right| <0.8$, where $
\theta _{\gamma }^{\eta }$ is the photon helicity angle in the $\eta$ rest
frame. The $\eta ^{\prime }\rightarrow \eta \left( \gamma \gamma \right) \pi
^{+}\pi ^{-}$ candidate uses the same $\eta$ selections, including the
$\pi^0$ veto. The mass of $K^0_S$ candidates
is taken at the closest distance of approach between positively and negatively charged
tracks.

\begin{table}[!htb]
\caption{Mass windows used in selection of intermediate particles.}
\begin{center}
\begin{tabular}{ll}
\hline\hline
MODE & Mass Range (GeV/c$^{2}$) \\ \hline
$J/\psi \rightarrow e^+e^-$ & $2.95<M_{e^{+}e^{-}}<3.14$ \\ 
$J/\psi \rightarrow \mu^+ \mu^- $ & $3.06<M\left( \mu ^{+}\mu ^{-}\right) <3.14$
\\ 
$\phi \rightarrow K^{+}K^{-}$ & $1.004<M\left( K^{+}K^{-}\right) <1.034$ \\ 
$K_{S}^{0}\rightarrow \pi ^{+}\pi ^{-}$ & $0.489<M\left( \pi ^{+}\pi
^{-}\right) <0.507$ \\ 
$\eta \rightarrow \gamma \gamma $ & $0.529<M\left( \gamma \gamma \right)
<0.565 $ \\ 
$\eta \rightarrow \pi ^{+}\pi ^{-}\pi ^{0}$ & $0.529<M\left( \pi ^{+}\pi
^{-}\pi ^{0}\right) <0.565$ \\ 
$\eta ^{\prime }\rightarrow \eta \pi ^{+}\pi ^{-}$ & $0.938<M\left( \eta \pi
^{+}\pi ^{-}\right) <0.978$ \\ 
$\pi ^{0}\rightarrow \gamma \gamma $ & 0.$120<M\left( \gamma \gamma \right)
<0.150$ \\ \hline
\end{tabular}
\end{center}
\label{table-mass}
\end{table}

An additional requirement is applied to separate and remove two-jet-like continuum events from
more spherical $B$ meson decays. The thrust direction of the $B$
meson candidate and thrust direction of the recoiling other tracks in the
event are calculated. Typically, $\theta _{T}$, the angle between these two
directions is uncorrelated for $B\overline{B}$ events
and peaked at $\cos \theta _{T}=\pm 1$ for continuum events. The thrust
angle requirement for the $B$ decays is $\left| \cos \theta _{T}\right| <0.8.$

The PID criteria are listed in Table~\ref{table-pid} mode by mode.
The PID requirements for some modes are slightly more stringent for background rejection.
\begin{table}[!htb]
\caption{Particle identification requirements for each decay mode.}
\begin{center}
\begin{tabular}{lllll}
\hline\hline
& $J/\psi \rightarrow e^+e^-$ & $J/\psi \rightarrow \mu^+ \mu^- $ & $\phi
\rightarrow K^{+}K^{-}$ & $3^{rd}K^{\pm }$ \\ \hline
$J/\psi \phi K^{+}$ & $VTE+LE$ & $VLM+LM$ & $VTK+LK$ & $NP$ \\ 
$J/\psi \phi K_{S}^{0}$ & $VTE+LE$ & $VLM+LM$ & $VTK+LK$ &  \\ 
$J/\psi \phi $ & $VTE+VTE$ & $TM+TM$ & $VTK+LK$ &  \\ 
$J/\psi \eta $ & $VTE+VTE$ & $TM+TM$ &  &  \\ 
$J/\psi \eta ^{\prime }$ & $VTE+VTE$ & $TM+TM$ &  &  \\ \hline
\end{tabular}
\end{center}
\label{table-pid}
\end{table}
\newpage
\subsubsection{$\protect\smallskip B^{0}\rightarrow J/\psi \phi $ Mode}
\begin{figure}[!htb]
\begin{center}
\includegraphics[height=7cm]{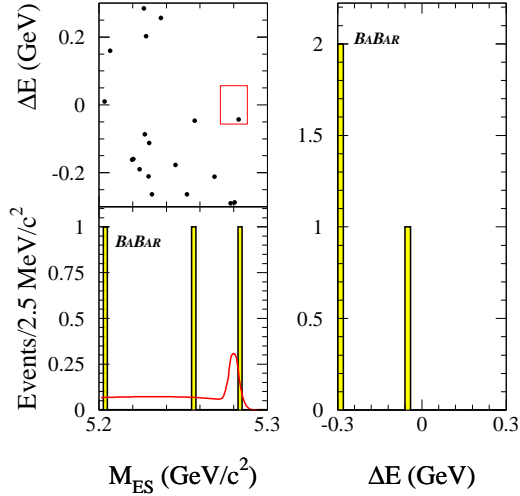}
\caption{
$\Delta E$ vs $M_{ES}$ (left top), $M_{ES}$ projection in $\Delta E$ signal region (left bottom), and $\Delta E$ projection in $M_{ES}$ signal region (right) for $B^{0}\rightarrow J/\psi \phi$.
}
\label{fig:jpub}
\end{center}
\end{figure}

This mode combines the $J/\psi $ and $\phi $ based on the selection described
in the previous section.  The resulting scatter plot of $\Delta E$ versus $M_{ES}$ is shown in Figure~\ref{fig:jpub} (left top).  The signal region is shown on the figure, and it is defined by $5.272<M_{ES}<5.288$ GeV/$c^{2}$ and $-0.057<\Delta E<0.057$ GeV.  The left bottom (right) plot shows the projection onto the $M_{ES}$ ($\Delta E$) axis for events that satisfy the $\Delta E$ ($M_{ES}$) requirement for the signal region. The curve overlaid on the $M_{ES}$ projection in this and the following figures is the sum of  an ARGUS function~\cite{argus-function} to model the combinatoric background and a Gaussian, where the Gaussian contains the background peaking in the signal region as well as the signal itself (see Section~\ref{ebu} for details). Statistically there is no significant signal for $B^{0}\rightarrow J/\psi \phi $. An upper limit on the branching fraction is described in the next section.

\subsubsection{$\protect\smallskip B\rightarrow J/\psi \phi K^{+}
$ and $J/\psi \phi K_{S}^{0}$ Modes}
\begin{figure}[!htb]
\begin{center}
\includegraphics[height=7cm]{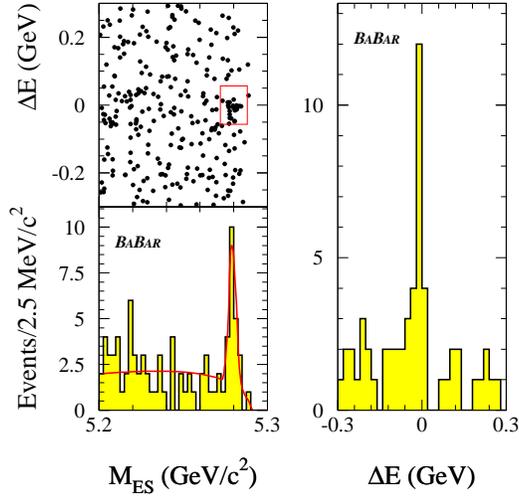}
\caption{
$\Delta E$ vs $M_{ES}$ (left top), $M_{ES}$ projection in $\Delta E$ signal region (left bottom), and $\Delta E$ projection in $M_{ES}$ signal region (right) for $B^{+}\rightarrow J/\psi \phi K^{+}$.
}
\label{fig:jpkub}
\end{center}
\end{figure}

\begin{figure}[!htb]
\begin{center}
\includegraphics[height=7cm]{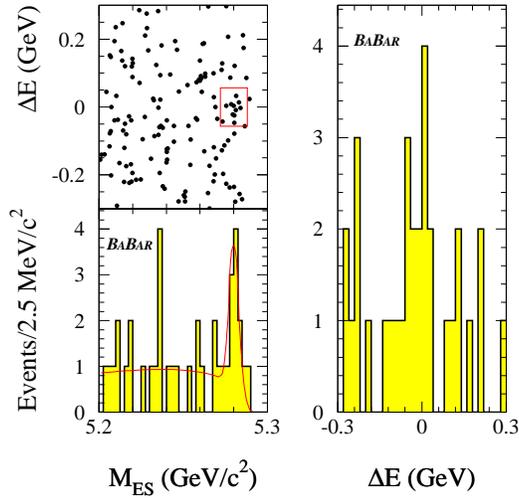}
\caption{
$\Delta E$ vs $M_{ES}$ (left top), $M_{ES}$ projection in $\Delta E$ signal region (left bottom), and $\Delta E$ projection in $M_{ES}$ signal region (right) for $B^{0}\rightarrow J/\psi \phi K_{S}^{0}$.
}
\label{fig:jpksub}
\end{center}
\end{figure}

In this mode we combine the $J/\psi$ and $\phi$ candidates described above with a charged kaon or $K^0_S$ candidate. The resulting scatter plot of $\Delta E$ versus $M_{ES}$ is shown in Figure~\ref{fig:jpkub} (left top) for $B^{+}\rightarrow J/\psi \phi K^+$.  The signal region is shown on the figure, and it is defined by $5.272<M_{ES}<5.288$ GeV/$c^{2}$ and $-0.057<\Delta E<0.057$ GeV.  The left bottom (right) plot shows the projection onto the  $M_{ES}$ ($\Delta E$) axis for events that satisfy the $\Delta E$ ($M_{ES}$) requirement for the signal region. The corresponding plots for $B^0\rightarrow J/\psi \phi K_{S}^{0}$ are shown in Figure~\ref{fig:jpksub}. The branching fractions are determined in the next section.


\subsubsection{$B^{0}\rightarrow J/\psi \eta $ Mode}

\begin{figure}[!htb]
\begin{center}
\includegraphics[height=7cm]{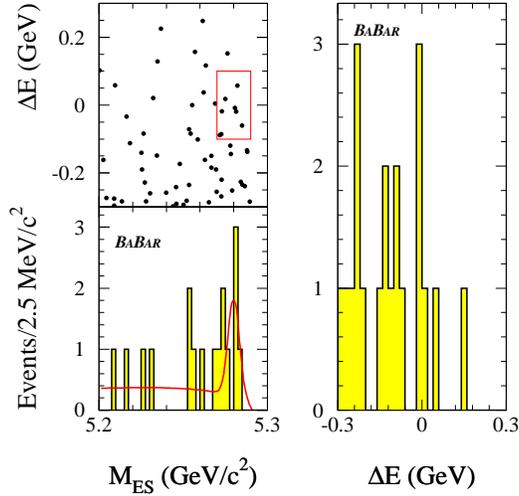}
\caption{
$\Delta E$ vs $M_{ES}$ (left top), $M_{ES}$ projection in $\Delta E$ signal region (left bottom), and $\Delta E$ projection in $M_{ES}$ signal region (right) for $B^{0}\rightarrow J/\psi  \eta(\gamma \gamma)$.
}
\label{fig:je2ub}
\end{center}
\end{figure}

\begin{figure}[!htb]
\begin{center}
\includegraphics[height=7cm]{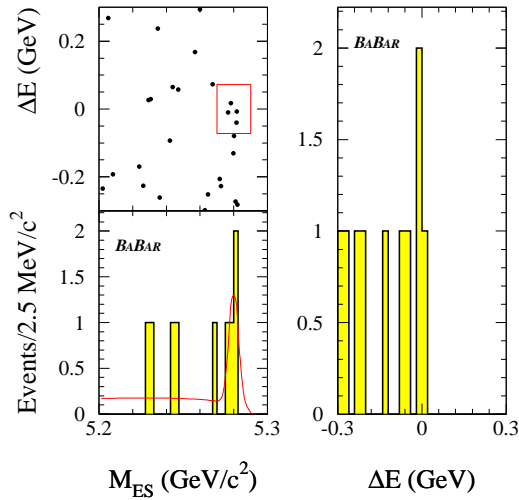}
\caption{
$\Delta E$ vs $M_{ES}$ (left top), $M_{ES}$ projection in $\Delta E$ signal region (left bottom), and $\Delta E$ projection in $M_{ES}$ signal region (right) for $B^{0}\rightarrow J/\psi \eta (\pi^+\pi^-\pi^0)$.
}
\label{fig:je3ub}
\end{center}
\end{figure}

For this mode, we combine a $J/\psi$ candidate with an $\eta$ candidate in the final states $\gamma \gamma $ or $\pi ^{+}\pi ^{-}\pi ^{0}$. The resulting scatter plot of $\Delta E$ versus $M_{ES}$ is shown in Figure~\ref{fig:je2ub} (left top) for the $\gamma\gamma$ mode and in Figure~\ref{fig:je3ub} (left top) for $\pi ^{+}\pi ^{-}\pi ^{0}$ mode. The left bottom plot and the right plot show the projections onto  $M_{ES}$ and $\Delta E$ respectively. The signal region is defined by $5.27<M_{ES}<5.29$~GeV/$c^{2}$ and $-0.1<\Delta E<0.1$ GeV for the $\gamma\gamma$ mode, and $5.27<M_{ES}<5.29$ GeV/$c^{2}$ and $-0.072<\Delta E<0.072$ GeV for $\pi ^{+}\pi ^{-}\pi ^{0}$ mode. No statistically significant signal is observed. Upper limits on the branching fractions for these modes are described in the next section.


\subsubsection{$B^{0}\rightarrow J/\psi \eta^{\prime } $ Mode}

\begin{figure}[!htb]
\begin{center}
\includegraphics[height=7cm]{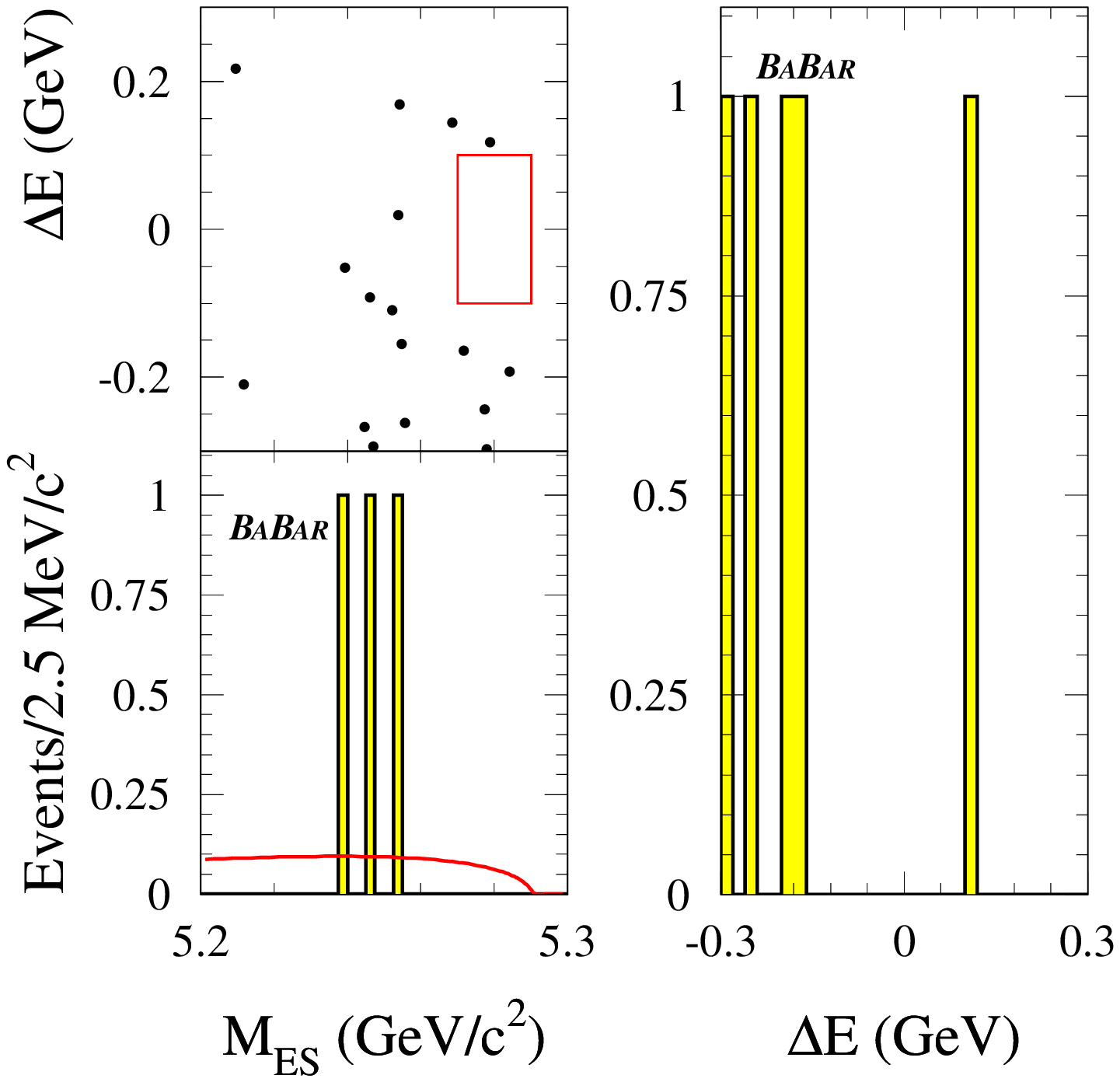}
\caption{
$\Delta E$ vs $M_{ES}$ (left top), $M_{ES}$ projection in $\Delta E$ signal region (left bottom), and $\Delta E$ projection in $M_{ES}$ signal region (right) for $B^{0}\rightarrow J/\psi \eta ^{\prime }$.
}
\label{fig:jepub}
\end{center}
\end{figure}

In this mode we combine  $J/\psi $ and $\eta ^{\prime }$ candidates.  The resulting scatter plot of $\Delta E$ versus $M_{ES}$ is shown in Figure~\ref{fig:jepub} (left top) for $B^{0}\rightarrow J/\psi \eta ^{\prime }$.  The signal region is defined by $5.27<M_{ES}<5.29$ GeV/$c^{2}$ and $-0.1<\Delta E<0.1$ GeV. The left bottom (right) plot shows the projected $M_{ES}$ ($\Delta E$) distribution for events that satisfy the $\Delta E$ ($M_{ES}$) requirement for the signal region. There is no significant evidence for $B^{0}\rightarrow J/\psi \eta ^{\prime }$. An upper limit on the branching fraction is determined in the next section.

\section{Efficiencies, Backgrounds and Systematic Uncertainties}
\label{ebu}
The efficiencies for each mode are determined by Monte Carlo simulation
where three-body phase space is assumed for the three body modes ($J/\psi
\phi K^{+},$ $J/\psi \phi K_{S}^{0})$, two-body phase space for the
vector-vector mode $\left( J/\psi \phi \right) $, and helicity amplitude
matrix elements for vector-pseudoscalar modes ($J/\psi \eta ,$ $J/\psi \eta
^{\prime })$. The statistical error due to the number of Monte Carlo events is included as part of the systematic error.

The background in the $M_{ES}$ distributions can be described by an
ARGUS function for the combinatoric background, plus a Gaussian function for the peaking background. The combinatoric background, denoted $N_{ARGUS}$, is due to continuum events, $B\overline{B}$ events with at least one $J/\psi$, and $B\overline{B}$ events without a $J/\psi$. The peaking background, denoted $N_{J/\psi-Gauss}$, comes only from $B\overline{B}$ events with a $J/\psi$. The shape of the ARGUS term is determined mode by mode by fitting an ARGUS function to the $M_{ES}$ distribution from a special set of events in the data where the $J/\psi$ is replaced by a fake $J/\psi$. The fake $J/\psi$ is selected with identical selection criteria in each mode except for logically reversing the lepton identification. This provides a large sample in each mode whose $M_{ES}$ distribution can be fitted and represents the ARGUS shape. The normalization of the combinatoric background for each mode is obtained from a fit to the $M_{ES}$ distributions in the $\Delta E$ signal region of the on-peak data. The integral of this function in the signal region is $N_{ARGUS}$. This method of determining $N_{ARGUS}$ has been checked with Monte Carlo simulation, off-peak data and $\Delta E$ and  $J/\psi$ mass sidebands from on-peak data. The peaking background $N_{J/\psi-Gauss}$ is determined from a sample of Monte Carlo $\B\Bb$ events that is normalized to the equivalent data integrated luminosity and contains at least one decay of $\jpsi\rightarrow$ leptons. The $M_{ES}$ distribution from this sample is fit with an ARGUS function and a Gaussian in the $\Delta E$ signal region where the normalizations are allowed to vary. The number of events in the resulting Gaussian fit is the Monte Carlo estimation of the peaking background $N_{J/\psi-Gauss}$. The sum of $N_{ARGUS}$ plus $N_{J/\psi-Gauss}$ gives $n_{b}$, the total number of background candidates in the signal region and its error, $\sigma _{b}$. The combinatoric background is by far the dominant background in all modes except the $B^{0}\rightarrow J/\psi \eta (\pi^+\pi^-\pi^0)$ mode, where the peaking component reaches $\sim20\%$ of the total background.

The following sources of systematic uncertainty are considered.

\begin{itemize}
\item  Uncertainty in the number of $B\overline{B}$ events (column labeled $\Delta N_{B\overline B}$ in Table~\ref{table-systematics}).

\item Uncertainty from secondary branching fractions (column labeled SBF in Table~\ref{table-systematics}).

\item  Monte Carlo statistical error (column labeled MC in Table~\ref{table-systematics}).

\item  Uncertainties in PID, tracking efficiency and photon detection efficiency (column labeled PidTrkG in Table~\ref{table-systematics}).

\item  Variations in the event selection criteria (column labeled EvtSel in Table~\ref{table-systematics}).

\item  Background parameterization (column labeled BkgdP in Table~\ref{table-systematics}).
\end{itemize}

The secondary branching fraction uncertainty combines all errors from the Particle Data Group (PDG)~\cite{pdg} for each mode. The fractional
uncertainty in $N_{B\overline{B}}$  is $1.6\%.$  The uncertainty from PID, tracking efficiency and photon detection efficiency is based on the study of the control samples. The uncertainty due to event selection includes varying all event selection criteria by a reasonable amount and determining the effect on the branching fraction. The uncertainty from background parameterization is estimated by using $\Delta E$ sideband information. The largest systematic error comes from varying the event selection criteria and no single variation dominates this systematic in any mode.

The total systematic error combines all these separate errors in quadrature
mode by mode. The individual systematic uncertainties are listed in Table~\ref
{table-systematics}.

\begin{table}[!htb]
\caption{Systematic error summary.}
\begin{center}
\begin{tabular}{llllllll}
\hline\hline
Mode &$\Delta N_{B\overline B}$ & SBF & MC & PidTrkG&EvtSel & BkgdP & Total ($\sigma_T$) \\ \hline
$J/\psi \phi $ &1.6\%& 2.2\% & 1.6\% & 6.7\% & 11.7\% & 12.0\% & 18.3\% \\ 
$J/\psi \phi K^+$ &1.6\%& 2.2\% & 1.6\% & 8.2\% & 11.5\% & 5.9\% & 15.6\% \\ 
$J/\psi \phi K_{S}^0$ &1.6\%& 2.2\% & 2.1\% & 8.3\% & 14.8\% & 1.9\% & 17.5\% \\ 
$J/\psi \eta \left( \gamma \gamma \right) $&1.6\% & 1.8\% & 1.6\% & 2.9\% & 14.3\%
& 6.9\% & 16.4\% \\ 
$J/\psi \eta \left( 3\pi \right) $&1.6\% & 2.4\% & 2.2\% & 7.7\% & 13.9\% & 8.0\%
& 16.5\% \\ 
$J/\psi \eta ^{\prime }$&1.6\% & 3.8\% & 4.6\% & 5.7\% & 11.7\% & 7.1\% & 16.1\%
\\ \hline
\end{tabular}
\end{center}
\label{table-systematics}
\end{table}

\section{Branching Fractions and Upper Limits}

The branching fraction determination uses a simple subtraction of events in the signal region. 
The number of signal events is $n_{s}=n_{0}-n_b$, 
where the term $n_{0}$ is  the number of data events in the signal region, and $n_b$ is  the total background described in Section~\ref{ebu}. 

The modes $J/\psi \phi K^{+}$ and $J/\psi \phi K_{S}^{0}$ have significant signals:
$J/\psi \phi K^{+}$ is 3.1 statistical standard deviations from zero, while $J/\psi \phi K_{S}^{0}$ is 2.7 statistical standard deviations from zero. The calculated branching fraction
is based on the Monte Carlo efficiency, $n_{s}$, $N_{B\overline{B}}$, and the secondary branching fractions for the $J/\psi ,$ $\phi $, and $
K_{S}^{0}$ from PDG~\cite{pdg}. The results are summarized in Table~\ref{signal-bf} including
the total summed background events in the signal region. The first error is the
statistical error, and the second error is the systematic error $\sigma_T$ taken
from Table~\ref{table-systematics}. The derived result for $B^0 \rightarrow J/\psi \phi K^0$ is also shown in Table~\ref{signal-bf}.
\begin{table}[!htb]
\caption{Branching fractions for $J/\psi \phi K^{+}$, $J/\psi \phi K_{S}^{0}$ and the derived result for  $J/\psi \phi K^0$.}
\begin{center}
\begin{tabular}{llllll}
\hline\hline
Mode & Efficiency &$n_0$& $n_{s}\pm\sigma(n_{s})$ & $n_b\pm\sigma_b$ &Branching Fraction \\ \hline
$J/\psi \phi K^{+}$ & $10.6\%$&23 & $15.2\pm 4.8$
& $7.8\pm 0.6$ &$
(~4.4\pm 1.4(stat)\pm 0.7(syst)){\times 10}^{-5}$ \\ 
$J/\psi \phi K_{S}^{0}$ & ${8.6\%}$&13 & ${9.7\pm 3.6}$
 & {${3.3\pm 0.4}$} & 
{$(~{5.1\pm 1.9(stat)\pm 0.9(syst)} ) {\times 10}^{-5}$} \\ 
$J/\psi \phi K^{0}$ & & &&& {$( {10.2\pm 3.8(stat)\pm 1.8(syst)} ) {\times 10}^{-5}$}\\
\hline
\end{tabular}
\end{center}
\label{signal-bf}
\end{table}

For the modes with no signal or weak statistical evidence ($J/\psi \phi $, $
J/\psi \eta $, $J/\psi \eta ^{\prime }$) an upper limit is set. The upper limit method uses the number of data events
counted in the signal region, $n_{0}$, $n_{b}$, and its error $\sigma_b$ (described in Section~\ref{ebu}), in the signal region and the total systematic uncertainty $\sigma _{T}\left( \%\right) $ from Table~\ref{table-systematics}. Once we obtain $n_{0}$, $n_{b}\pm \sigma _{b}$, and $\sigma _{T},$
then we assume these two uncertainties ($\sigma _{b},\sigma _{T}$) are
uncorrelated and Gaussian, the upper limit $N_{90\%}$ is obtained by
folding the Poisson distribution with two normal distributions for
these two uncertainties and integrating it to the 90\% confidence level. We list the variables in Table~\ref{table-upperlimit-bf} to obtain, $N_{90\%}$,
the number of events for a 90\% upper confidence limit. Then using the upper limit $N_{90\%}$, the efficiency and $N_{B\overline{B}}$, we determine the resulting upper
limits on the branching fractions which are also shown in Table~\ref{table-upperlimit-bf}.

\begin{table}[!htb]
\caption{90\% upper confidence limits.}
\begin{center}
\begin{tabular}{lllllll}
\hline\hline
Mode & Efficiency& $n_{0}$ & $n_{b}\pm \sigma _{b}$ & $\sigma _{T}\left( \%\right) $ & $
N_{90\%}$ &90\% C.L. Upper Limit \\ \hline
$B\rightarrow J/\psi \phi $ &${12.1\%}$& ${1}$ & ${0.3\pm 0.2}$ & ${18.3}$ & ${3.70}$ & ${<.95\times 10^{-5}}$\\
$B\rightarrow J/\psi \eta \left( \gamma \gamma \right) $ & ${15.5\%}$ &${8}$ & ${1.7\pm 0.4}$ & ${16.4}$ & {${11.8}$} & ${<3.0}\times 10^{-5}$\\ 
$B\rightarrow J/\psi \eta \left( \pi^+\pi^-\pi^0 \right) $ &${8.7\%}$ & ${4}$ & ${1.5\pm 0.9}$ & ${16.5}$ & {${6.86}$}& ${<5.2}\times 10^{-5}$\\ 
$B\rightarrow J/\psi \eta$ combined & &&&&& ${<2.7}\times 10^{-5}$ \\
$B\rightarrow J/\psi \eta ^{\prime }$ & ${2.5\%}$& ${0}$ &${0.5\pm 0.3}$ & ${16.1}$ & {${1.84}$} &${<6.4}\times 10^{-5}$ \\ \hline
\end{tabular}
\end{center}
\label{table-upperlimit-bf}
\end{table}

\section{Conclusions}

We observe evidence for $B\rightarrow J/\psi \phi K$ in two
modes and determine the branching fractions ${\cal B}$($B\rightarrow J/\psi \phi K^{+}$)=
{${(4.4\pm 1.4(stat)\pm 0.7(syst))\times 10}^{-5}$} and ${\cal B}$($B\rightarrow J/\psi
\phi K_{S}^{0}$)={$(5.1\pm 1.9(stat)\pm 0.9(syst))\times 10^{-5}$}. The branching fraction for $
B\rightarrow J/\psi \phi K$ is consistent with CLEO results~\cite{cleo}. Upper
limits have been determined for the modes $B\rightarrow J/\psi \phi ,$ $
J/\psi \eta $, and $J/\psi \eta ^{\prime }$. However, the two $B\rightarrow
J/\psi \eta$ upper limits in Table~\ref{table-upperlimit-bf} would correspond to a combined branching fraction of $(1.6\pm0.6(stat.)\pm0.2(syst.))\times 10^{-5}$, which is comparable to the $
B\rightarrow J/\psi \pi ^{0}$ branching fraction.

\section{Acknowledgments}
\label{sec:Acknowledgments}
We are grateful for the 
extraordinary contributions of our \pep2\ colleagues in
achieving the excellent luminosity and machine conditions
that have made this work possible.
The success of this project also relies critically on the 
expertise and dedication of the computing organizations that 
support \babar.
The collaborating institutions wish to thank 
SLAC for its support and the kind hospitality extended to them. 
This work is supported by the
US Department of Energy
and National Science Foundation, the
Natural Sciences and Engineering Research Council (Canada),
Institute of High Energy Physics (China), the
Commissariat \`a l'Energie Atomique and
Institut National de Physique Nucl\'eaire et de Physique des Particules
(France), the
Bundesministerium f\"ur Bildung und Forschung
(Germany), the
Istituto Nazionale di Fisica Nucleare (Italy),
the Research Council of Norway, the
Ministry of Science and Technology of the Russian Federation, and the
Particle Physics and Astronomy Research Council (United Kingdom). 
Individuals have received support from 
the A. P. Sloan Foundation, 
the Research Corporation,
and the Alexander von Humboldt Foundation.

$\smallskip $


\begin{thebibliography}{99}
\bibitem{babar-charmonium}M. Alam $et\ al.$ [CLEO collaboration], Phys. Rev. \textbf{D34}, 3279 (1986);\\
B. Aubert $et\ al.$ [$\babar$ collaboration], Phys. Rev. \textbf{D65},
32001 (2002).

\bibitem{babar-jpsirho}M. Bishai $et\ al.$ [CLEO Collaboration], Phys. Lett. \textbf{B369}, 186 (1996);\\
 $\babar$ Collaboration,
Measurement of the Branching Fraction
$B^0\rightarrow J/\psi \pi\pi$, presented to this conference.

\bibitem{resc} A. Dighe, I. Dunietz, R. Fleischer, Phys. Lett. \textbf{B433}, 147 (1998);\\
M. Suzuki, Phys. Rev. \textbf{D64}, 117503 (2001).

\bibitem{deandrea}  A. Deandrea \textit{et al.}, Phys. Lett. \textbf{B318}, 549 (1993).

\bibitem{jpsi-eta-br}  We use ${\cal B}(B^{0}\rightarrow J/\psi \pi
^{0})=\left( 2.0\pm 0.6\pm 0.2\right) \times 10^{-5}$ from the $\babar$
Collaboration, Phys. Rev. \textbf{D65}, 32001 (2002).

\bibitem{L3}  M. Acciarri $et\ al.$ [L3 Collaboration], Phys. Lett. \textbf{B391}, 481 (1997).

\bibitem{cleo}A. Anastassov $et\ al.$ [CLEO Collaboration], Phys. Rev. Lett. \textbf{84}, 1393 (2000).

\bibitem{close}  F. E. Close, I. Dunietz, P.R. Page, S. Veseli and H.
Yamamoto, Phys. Rev. \textbf{D57}, 5653 (1998).

\bibitem{babar-det}  B. Aubert $et\ al.$ [$\babar$ Collaboration], Nucl. Instr. and Methods \textbf{A479}, 1 (2002).

\bibitem{LAT}  A. Drescher \textit{et al.}, Nucl. Instr. and Methods \textbf{
A237}, 464 (1985).

\bibitem{zernike}  Ralph Sinkus and Thomas Voss, Nucl. Instr. and Methods \textbf{A391}, 360 (1997).

\bibitem{argus-function}  H. Albrecht \textit{et al.} [ARGUS Collaboration], Z. Phys \textbf{C48},
543 (1990).

\bibitem{pdg}  D.E. Groom \textit{et al.} [Particle Data Group], Eur. Phys. J. C. \textbf{15},
1 (2000).


\end{thebibliography}
\end{document}